\documentclass[11pt,oneside]{article}                                                                            
\usepackage[english]{babel}
\usepackage[utf8]{inputenc}
\usepackage{enumitem}
\usepackage{etaremune}
\usepackage{calc}
\usepackage{amssymb}
\usepackage{amsmath}
\usepackage[all]{xy}
\usepackage{bm}
\usepackage{latexsym}
\usepackage[dvips]{graphicx}
\usepackage{tikz}
\usepackage{longtable}
\usepackage{booktabs}
\usepackage{colortbl}
\usepackage{diagbox}
\usepackage{psfrag}
\usepackage{rotating}
\usepackage{lscape}
\usepackage{color}
\usepackage{multicol}
\usepackage{multirow}
\usepackage{hhline}
\usepackage{cancel}
\usepackage{hyperref}
\usepackage{endnotes}
\usepackage{tikz}
\usepackage[firstabbrev]{authorindex}
\usepackage[subnum]{cases}
\usepackage[numbers,square,sort&compress]{natbib}
\let\baraccent=\= 
\renewcommand{\=}[1]{\stackrel{#1}{=}} 

\begin{document}
\pagestyle{plain}
\begin{center}
\Large{A Predictive Strategy for the Iterated Prisoner's Dilemma}\vspace*{0.5cm}\\
\normalsize{Robert Prentner\\ 
\href{mailto:robert.prentner@gmail.com}{robert.prentner@gmail.com}
}
\end{center}
%
\begin{center}
\subsection*{Abstract}
\end{center}
The iterated prisoner's dilemma is a game that produces many counter-intuitive and complex behaviors in a social environment, based on very simple basic rules. It illustrates that cooperation can be a good thing even in a competitive world, that individual fitness needs not to be the most important criteria of success, and that some strategies are very strong in a direct confrontation but could still perform poorly on average or are evolutionarily unstable.

In this contribution, we present a strategy -- \textsf{PREDICTOR} -- which appears to be ``sentient'' and chooses to cooperate when playing against some strategies, but defects when playing against others, without the need to record ``tags'' for its opponents or an involved decision-making mechanism. To be able to operate in the highly-contextual environment, as modeled by the iterated prisoner's dilemma, \textsf{PREDICTOR} learns from its experience to choose optimal actions by modeling its opponent and predicting a (fictive) future. 

It is shown that \textsf{PREDICTOR} is an efficient strategy for playing the iterated prisoner's dilemma and is simple to implement. In a simulated and representative tournament, it achieves high average scores and wins the tournament for various parameter settings. \textsf{PREDICTOR} thereby relies on a brief phase of exploration to improve its model, and it can evolve morality from intrinsically selfish behavior.
\\\\
\newpage

\section{Introduction}
\label{sec:intro}
The iterated prisoner’s dilemma (\textsf{IPD}) has been called “the E.coli of social psychology” \citep{Axe84}. Though simple in structure, it illustrates many intriguing features of social interaction among agents:

\begin{itemize}
\item The single-shot version of the game (the ``stage-game'') has a unique Nash-equilibrium, with the consequence that any game played by two rational players would end in mutual defection. But the same outcome would not necessarily obtain in the iterated version of the game, given that the stage-game is iterated infinitely often or that players have only limited knowledge about the number of turns played.
\item During each turn, each player has a choice of two possible actions: cooperate ($C$) or defect ($D$). The game is described by a $2 \times 2$-payoff matrix that associates payoffs to each player at each turn. The individual payoffs are not only dependent on a player's own choice but also the choice of the other player.  
\item The \textsf{IPD} was originally studied to understand the evolution and prevalence  of cooperation among agents. Both blind cooperation and blind defection would lead to low payoffs in the iterated game (even though the latter is the ``rational'' choice in the single-shot game). 
The payoff matrix further makes sure that repeated cooperation between players leads to an overall better payoff than alternating between cooperation and defection ($2\times PO(CC) > PO(CD) + PO(DC)$). 
Cooperation can be beneficial in some circumstance but not others (where it is better to defect). This depends on the long-term strategy chosen by the other player. 

For example, given an opponent who chooses to always defect, it is the best choice to also defect on every turn. The same is true when playing against a player who always cooperates. The worst outcome is to always cooperate against an opponent who always defects. However, when playing against an opponent who chooses to cooperate after previous cooperation and to defect otherwise, it is best to always cooperate. 
The \textsf{IPD} thus illustrates that a strategy that is sub-optimal on the short run can be a very good strategy on the long run. 
\end{itemize}
Previous work on the \textsf{IPD} focused mostly on either of two things: how to devise (e.g. hard-code) a strategy that performs good in a round-robin style tournament \cite{Axe80a,Axe80b,Ste12,Har17} and how to adapt  the \textsf{IPD} to model quasi-biological evolution \cite{Axe88,Now93,Gar18}. 

It was found through simulations that particular strategies 
perform surprisingly well in the game. It is important to note that performing well in a tournament does not require to win many individual games. It simply means that a strategy received higher payoffs \textit{on average} across all matches: If player one lost all its games by a narrow margin but still reaped reasonably high payoffs each time, and players two and three performed badly when playing against each other, player one would win the tournament even though it lost all of its individual matches (Table \ref{tab:abc}). 
\begin{table}[hbt]
\centering
\label{tab:abc}
\caption{Winning is not everything. A fictional game between three players where the best-ranked player has not won any game in the contest and the player with most wins has the worst cumulative score.}
\begin{tabular}{llllll}
\\\toprule
 & player 1 & player 2 & player 3 & cumulative score & nr. wins\\\hline
player 1 & 2.25	& 2.5 &	2.25	& 	7.0 & 0\\
player 2 & 2.75	&3.	&1. &	 6.75	 &1\\
player 3 & 2.5	& 1.5 & 	1.75	& 5.75	& 2             \\ \bottomrule
\end{tabular}

\end{table}

Based on Axelrod \cite{Axe84}, one could classify successful strategies into ``nice'', ``fair'' and ``forgiving”, meaning that they are willing to cooperate, but do not do so blindly and retaliate with defection when playing against a strategy that previously defected. Moreover, successful strategies tend to disregard previous defection after some time. The perhaps best-known example of such a strategy is Tit-for-Tat (\textsf{TFT}), which is a surprisingly simple strategy that takes only into account what the opponent played on the previous turn. \textsf{TFT} won the first two tournaments of Axelrod \cite{Axe80a,Axe80b}, even though participants in the tournament were previously aware of its existence. Most likely, the simplicity of \textsf{TFT} triggered the belief that a more complex strategy would easily do better than \textsf{TFT} in the tournament. However, this was not the case. 

Sometimes it pays off to play a simple strategy, in particular in more realistic scenarios where simple and fast strategies have the advantage of being less costly and resource-intensive \cite{Kah82,Pea83,Gig91,Kah11}.
But in contrast to folklore about the \textsf{IPD}, it is not necessary to play a simple strategy in order to win a tournament. In general, and with the aid of computational optimization techniques, it is possible to design a strategy that is complex and which will beat many of the much simpler strategies in the game. 

Also spatial distribution, determining e.g. how often two strategies compete against each other, has potentially a huge impact on the overall results. This is of particular relevance for evolutionary extensions and consistent with the approach of ``evolutionary graph theory'' \cite{Lie05,Pav18}.

Furthermore, strategies which performed well in one particular tournament need not perform well in a different tournament. This depends, consistent with the general thrust of the \textsf{IPD}, precisely on what other strategies they are pitted against. For example, while \textsf{TFT} was the winning strategy in both of Axelrod's initial tournaments, in different settings \textsf{TFT} would not win. Such was the case in the 2012 tournament of Stuart \& Plotkin \cite{Ste12}, where \textsf{TFT} came in third (not first). This could be easily checked by running different tournaments with different combatants which is now possible by using predefined libraries (e.g the Axelrod Python library \cite{Axe450}). 

A particular kind of strategy was discovered by Press \& Dyson in 2012 \cite{Pre12}  through the application of some rather straightforward mathematical tools from probability theory and linear algebra. 
They made a series of observations: (i) any strategy with finite (but potentially large) memory could be exhaustively represented in terms of a Markovian matrix that assigns the probability of choosing to cooperate in response to each previous state of the game’s history, (ii) for an \textit{indefinitely repeated} game, strategies with a large memory do not have any advantage when paired against a strategy which would only consider the results of the previous round in the game (so called ``memory-1 strategies''), and (iii) there exist memory-1 strategies which could either set the payoff of their opponent to some fixed value or enforce a linear relation between the opponent's payoff and their own payoff, effectively extorting the other player. Such strategies were called ``zero-determinant'' (\textsf{ZD}) strategies, because their derivation included setting the determinant of a $4 \times 4$ matrix to zero. This determinant is a function of the desired linear relation between payoffs and the probabilities characterizing a strategy and thus guarantees desired payoff relations.

In a subsequent tournament, Stewart \& Plotkin \cite{Ste12} demonstrated that a particular \textsf{ZD}-strategy (\textsf{ZDGTFT-2}) won the contest, while another \textsf{ZD}-strategy (\textsf{EXTORT-2}) won most of its matches (beaten only by the trivial ``All-Defect” strategy which is unbeatable in principle). Both strategies enforce a linear relationship between payoffs such that the strategy receives twice the payoff of its opponent above a threshold. In the case of \textsf{ZDGTFT-2} this results in:  $P_x=2\cdot P_y - 3$; for \textsf{EXTORT-2} this results in $P_x=2\cdot P_y-1$.  

While \textsf{EXTORT-2} might seem to be the ``better'' strategy at first sight, it does not perform well overall. Why is this so? As one could infer from the above equations, how such a strategy effectively performs depends on the opponent's payoff. While \textsf{EXTORT-2} grants less of the joint share to its opponent, it might still perform worse overall if the opponent's payoffs are low. A more generous strategy such as \textsf{ZDGTFT-2} might thus profit from a better performing opponent even though it grants them a comparatively larger fraction of the joint payoff to its opponent. 

However, it was soon pointed out that \textsf{ZD}-strategies are not evolutionary stable; even though a \textsf{ZD}-strategy could enforce that the maximal payoff of \textit{its} opponent is always smaller or equal to its own (and thus that it will win many individual matches), it might still be beaten on average by other strategies which could reap more payoff values in other matches (compare again Tab. 1). While \textsf{ZD}-strategies can control the payoff in any match they take part in, they (obviously) cannot control matches from which they are absent. Another crucial issue stems from self-play where a \textsf{ZD}-strategy is pitted against itself. If such a strategy plays well against other strategies, it might initially increase its proportion in a population, but it will eventually face a drawback because it will have to compete against its own offspring fairly quickly \cite{Hin13}. Mutation and spatial structure might even worsen this effect.

It is known that the \textsf{IPD} illustrates that cooperation can be a good thing, that individual fitness need not be the most import criteria, and that some strategies are very strong in a direct confrontation but could still perform poorly on average or are evolutionarily unstable.
But it is suggestive to also believe that players which could sufficiently ``look ahead into the future'' and estimate the strategy of the other players are at an advantage in the game. Indeed, one consequence of Press \& Dyson's \cite{Pre12} analysis has been under-appreciated.  Press \& Dyson note that ``any strategy which will blindly adapt'' (i.e. only try to optimize its payoffs) is doomed to being extorted. Only players which are, in their words, ``sentient'', i.e. players which have an idea of what its opponent will play and then make a judgment of whether it is ok to conform to the extortion or not (in which case it is better to defect), can force an extortioner to change its strategy to a fairer one.

It is therefore mandated to construct a strategy that does exactly this: predict the most likely behavior of its opponent and then select an appropriate action. Our approach differs from recent work on the \textsf{IPD} in that it is neither the goal to find the ``best'' strategy for a particular tournament, e.g. by applying an optimization  algorithm to one of the many proposed architectures for designing strategies \cite{Har17}, or to understand the evolution of ``islands'' of cooperation within a population \cite{Gar18}. On the contrary, we believe that it is a hallmark of intelligent strategies to cooperate when paired with some strategies, but to defect when paired against others.

However, exactly when and why to choose whether to defect or cooperate is difficult to estimate a priori. It involves prediction and decision in a social (and uncertain) setting. The \textsf{IPD} might help to illustrate many tenets of social intelligence and learning though experience: in particular model-based learning,  the trade-off between``exploration'' and ``exploitation'', and the evolution of moral-looking behavior.
\section{Model}
\label{sec:model}

We use the standard payoffs of the \textsf{IPD}, where mutual defection ($DD$) is slightly more advantageous than cooperation when the other is defecting ($CD$) but less advantageous than either mutual cooperation ($CC$) or defection when the other is cooperating ($DC$). In short: $PO(CD) < PO(DD) < PO(CC) < PO(DC)$. The ``classical'' payoff matrix is depicted in Table \ref{tab:payoff}.
\begin{table}[hbt]
\label{tab:payoff}
\centering
\caption{Standard payoffs for player 1 in the \textsf{IPD}; payoffs for player 2 are symmetric. Note that defection is always better than cooperation in case the other decision is known. However, mutual cooperation is better than mutual defection.}
\begin{tabular}{llll}
\\\toprule
& & $C$  & $D$ \\\hline
$PO(\cdot\,\cdot)$ & $C$ & $3$	& $0$\\
$PO(\cdot\,\cdot)$  & $D$ & $5$	& $1$ \\ \bottomrule
\end{tabular}
\end{table}

Following Press \& Dyson \cite{Pre12}, we focus on memory-1 strategies and predictive strategies that compete against memory-1 strategies. Note that most strategies which have been successful in previous tournaments could be represented as memory-1 strategies, including Tit-for-Tat (\textsf{TFT}) \cite{Axe80a}, Win-stay-loose-shift (\textsf{WSLS}) \cite{Now93} and \textsf{ZD}-strategies \cite{Pre12}. There is in principle no obstacle to generalize the presented method to strategies other than memory-1 strategies or even to finite state machines which are known to be dense with respect to all existing (possibly non-computable) strategies in the \textsf{IPD} (cf. section 4.1. in \cite{Gar16}).

A memory-1 strategy is fully characterized by a four-vector of conditional probabilities, $p\big( A_t = C | A_{t-1} A'_{t-1}\big)$, each denoting the probability that cooperation ($A_t = C$) will be chosen, given that the player experienced his own previous decision and that of the other player ($A_{t-1}$ and $A'_{t-1}$ respectively). The probability of defecting ($A_t = D$) is then given by $p (A_t = D) =1-p (A_t=C)$. Each $A_{t-1}$ can be either $C$ or $D$, giving rise to $2^2$ possible combinations.

For example, \textsf{TFT} could be represented by the four-vector: 
\begin{equation}
\begin{pmatrix}
p(C|CC)= 1. \\
p(C|CD)=0.\\
p(C|DC) = 1.\\
p(C|DD)= 0. 
\end{pmatrix}.
\end{equation}
All four-vectors representing the strategies investigated in this work are listed in the \textcolor{blue}{supporting information S1}. Note also that this representation of strategies is equivalent to a Markov Decision Process.

Based on this, we implemented a strategy which predicts the other's decision and learns from its experience (\textsf{PREDICTOR}). Its own decisions affects (i) present payoffs, but also (ii) future payoffs. It thus estimates at each turn, based on its model (the four-vector) of what the other would do, all possible outcomes of the game for $n$ turns (for memory-1 strategies, only $n=2$ are needed). It then chooses the outcome with the highest expected payoff. We assume that \textsf{PREDICTOR} chooses the best possible action given it's limited knowledge (complete knowledge of the payoff matrix and partial knowledge of the other, as encoded in \textsf{PREDICTOR's} model). 
%
%

Given these assumptions, \textsf{PREDICTOR} chooses the action that maximizes expected payoff, according to:
\begin{equation}
A_1 = 
\begin{cases}
C\ \textrm{if}\ \langle PO \rangle_1 > \langle PO \rangle_3\\
D\  \textrm{if}\ \langle PO \rangle_1 \le \langle PO \rangle_3
\end{cases},
\end{equation}
with
\begin{eqnarray}
\langle PO \rangle_1 &=& PO(CC) p(C|X_0) +  PO(CD) p(D|X_0)\\
&&+\,PO (DC) \left[ p(C|CC)p(C|X_0) + p(C|CD)p(D|X_0)\right]\\
&&+\,PO (DD) \left[ p(D|CC)p(C|X_0) + p(D|CD)p(D|X_0)\right],
\end{eqnarray}
and:
\begin{eqnarray}
\langle PO \rangle_3 &=& PO(DC) p(C|X_0) +  PO(DD) p(D|X_0)\\
&&+\,PO (DC) \left[ p(C|DC)p(C|X_0) + p(C|DD)p(D|X_0)\right]\\
&&+\,PO (DD) \left[ p(D|DC)p(C|X_0) + p(D|DD)p(D|X_0)\right].
\end{eqnarray}
(For a derivation see the \textcolor{blue}{supporting information S2}). In a nutshell, PREDICTOR chooses to cooperate in case cooperation yields higher expected payoff after two turns, otherwise it chooses to defect. %

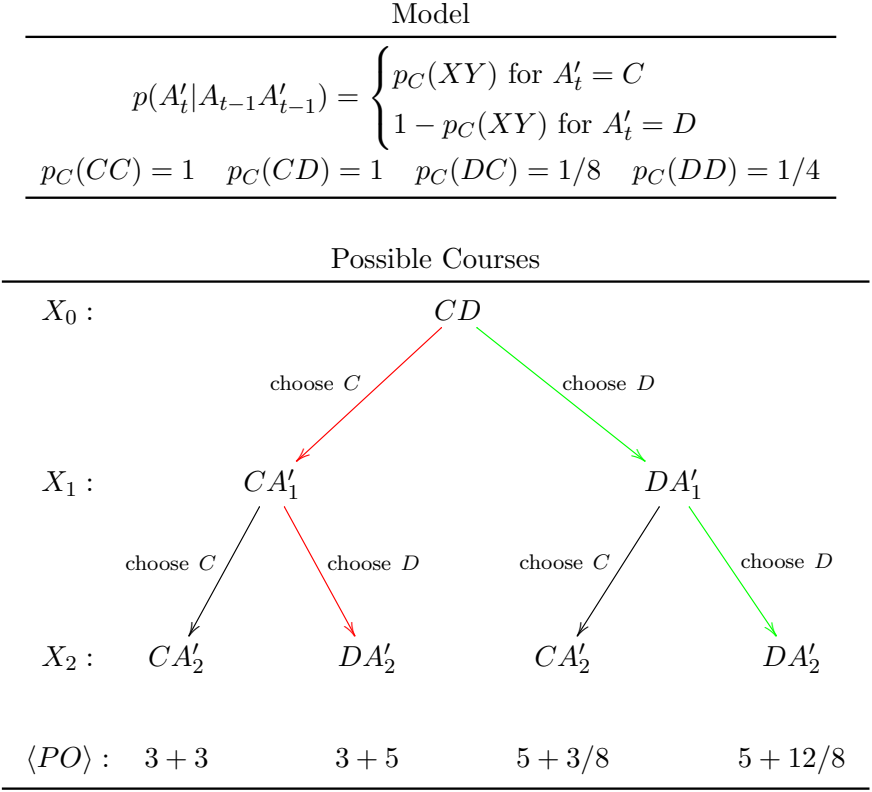
\begin{figure}[]
\centering
\begin{tabular}{llll}
\multicolumn{4}{c}{Model}\\
\toprule
\multicolumn{4}{c}{ $ p(A'_t| A_{t-1} A'_{t-1}) = \begin{cases} p_{C}(XY)\ \textrm{for}\ A'_t = C \\ 1-p_C(XY)\ \textrm{for}\ A'_t=D \end{cases}$}\\
$p_C(CC) = 1$  & $p_C(CD) = 1$ & $p_C(DC) = 1/8 $ & $p_C(DD) = 1/4$\\
\bottomrule
\end{tabular}
\vspace{0.5cm}\\
\begin{tabular}{c}
Possible Courses\\
\toprule
$
\xymatrixrowsep{0.75cm}
\xymatrixcolsep{0.25cm}
\xymatrix{
X_0: &&&& CD \ar@[red][ddll]_{\textrm{choose}\ C}\ar@[green][ddrr] ^{\textrm{choose}\ D}\\\\
X_1: && CA'_1 \ar[ddl]_{\textrm{choose}\ C} \ar@[red][ddr]^{\textrm{choose}\ D} &&&& DA'_1 \ar[ddl]_{\textrm{choose}\ C}\ar@[green][ddr]^{\textrm{choose}\ D}\\\\
X_2: & CA'_2 && DA'_2 & & CA'_2 && DA'_2
\\
\langle PO \rangle: & 3 + 3 && 3 + 5 && 5 + 3/8 && 5 + 12/8
} $\\
\bottomrule
\end{tabular}
\caption{Example of predicted payoff for $2$ turns. Top: \textsf{PREDICTOR}'s model of the opponent's strategy, which corresponds to the \textsf{ZD-GTFT-2} strategy. Bottom: The course of possible actions with total expected payoff. The course which corresponds to the actions $A_1 = C \rightarrow A_2 = D$ is highlighted in red; the course which corresponds to the actions $A_1 = D \rightarrow A_2 = D$ in green. Maximum expected payoff, relative to \textsf{PREDICTOR}'s model, is given by the red course. Accordingly, \textsf{PREDICTOR}'s final decision is to cooperate ($A_1 = C$). This is only so because \textsf{PREDICTOR} looks into the future. If only the current turn would be evaluated, the expected payoff would be larger for defection ($A_1 = D$).}
\label{fig:model}
\end{figure}

A simple example should illustrate this for a fictive round in Fig. \ref{fig:model}. Suppose that the previous decision was $X_0 = A_0 A'_0 = CD$. 
According to \textsf{PREDICTOR}'s model (Fig. \ref{fig:model}, top), the probability that the opponent will cooperate in the present round ($n=1$) is $1$ (i.e. $A'_1 = C$) 
If \textsf{PREDICTOR} countered by cooperation ($A_1 = C$) it would reap a payoff of $3$. On the next turn, the opponent would again cooperate with certainty, thus giving a total possible payoff of $3 + 5 = 8$ (\textsf{PREDICTOR} would chose to defect on the last, fictive, turn $A_2 =D$).\\
By contrast, if it initially chose to defect ($A_1 = D$), the payoff would be $5$ on the first turn. But this would also affect the probability of cooperation in the next ($n=2$) round, which, according to the model, would decrease to $p_C(DC) = 1/8$. Accordingly, the probability that the opponent defects is $p_D(DC) = 7/8$. If \textsf{PREDICTOR} finally chose to defect, it would thus reap a payoff of $5/8 + 7/8$ . In total, the expected payoff would be $6.5$. \\
The rational choice would thus be to choose cooperation in the present round to steer the game accordingly. 
%


\textsf{PREDICTOR} then updates its model, assuming a prior of $p(C|A_{t-1} A'_{t-1})=1/N_0$, by a simple mechanism: Update the prior probability to $2/(N_0+1)$ if the opponent's choice is cooperating and to $1/(N_0+1)$ in case it is defecting. Initially, ie. before the match, $N_0 = 2$, so the strategy starts with a maximally ignorant model. The model is set back to maximal uncertainty after any match is finished.
A \textcolor{blue}{flow-chart in the supporting information S3} further illustrates this. 

Furthermore, note that \textsf{PREDICTOR} \textit{does not} require to explicitly represent encounters previous to the adjacent past turn, i.e. it does not require to represent the history of a game more than a memory-1 strategies does. It just needs to be able to update its internal model at every step. 

To test the model, we implemented a round robin tournament featuring all memory-1 strategies from the previously reported tournament of Stewart \& Plotkin \cite{Ste12}, which also featured \textsf{ZD}-strategies, playing against \textsf{PREDICTOR}. This includes most of the ``classic'' strategies such as Tit-for-Tat (\textsf{TFT}), Win-Stay-Lose-Shift (\textsf{WSLS}), or Generous-Tit-for-Tat (\textsf{GTFT}). Note that average payoffs are heavily influenced by the selection of strategies for the tournament. We chose this setting for better comparison and to avoid further complications by introducing non-memory-1 strategies (although \textsf{PREDICTOR} could also be used here). 
Furthermore, even the order of matches might play a slight role. For this reason, the list of players has been randomized before initializing the tournament. 
Parameters were set to five iterations per pairing, $N_\textrm{iter} = 5$, with $N_\textrm{turns} = 200$ each. (These are the original setup chosen by Axelrod for his first tournament \cite{Axe80a}). During an initial exploration phase of duration $\Delta = p_\textrm{exp} N_\textrm{turn}$, \textsf{PREDICTOR} will choose a random action instead of ``exploiting'' its internal  prediction to improve its internal model. We recorded several tournaments for different values of $p_\textrm{exp}$ to test the effects of exploration.

We also tested the effect of randomizing an opponent's initial decision when playing against \textsf{PREDICTOR} (so, e.g., \textsf{TFT} would not always start with cooperation when playing against \textsf{PREDICTOR} but only when playing against other strategies in the tournament.) This decreases \textsf{PREDICTOR}'s average payoff but does not change the overall structure of our results.

\section{Results}
\label{sec:results}

\subsection{Round Robin}
Table \ref{tab:res} summarizes the results for one tournament with parameters $N_\textrm{iter} = 5$, $N_\textrm{turn} = 200$, $p_\textrm{exp} = 0.1$. The full payoff-matrix can be found in the \textcolor{blue}{supporting information S4}. 

\begin{figure}
\includegraphics[scale=0.76]{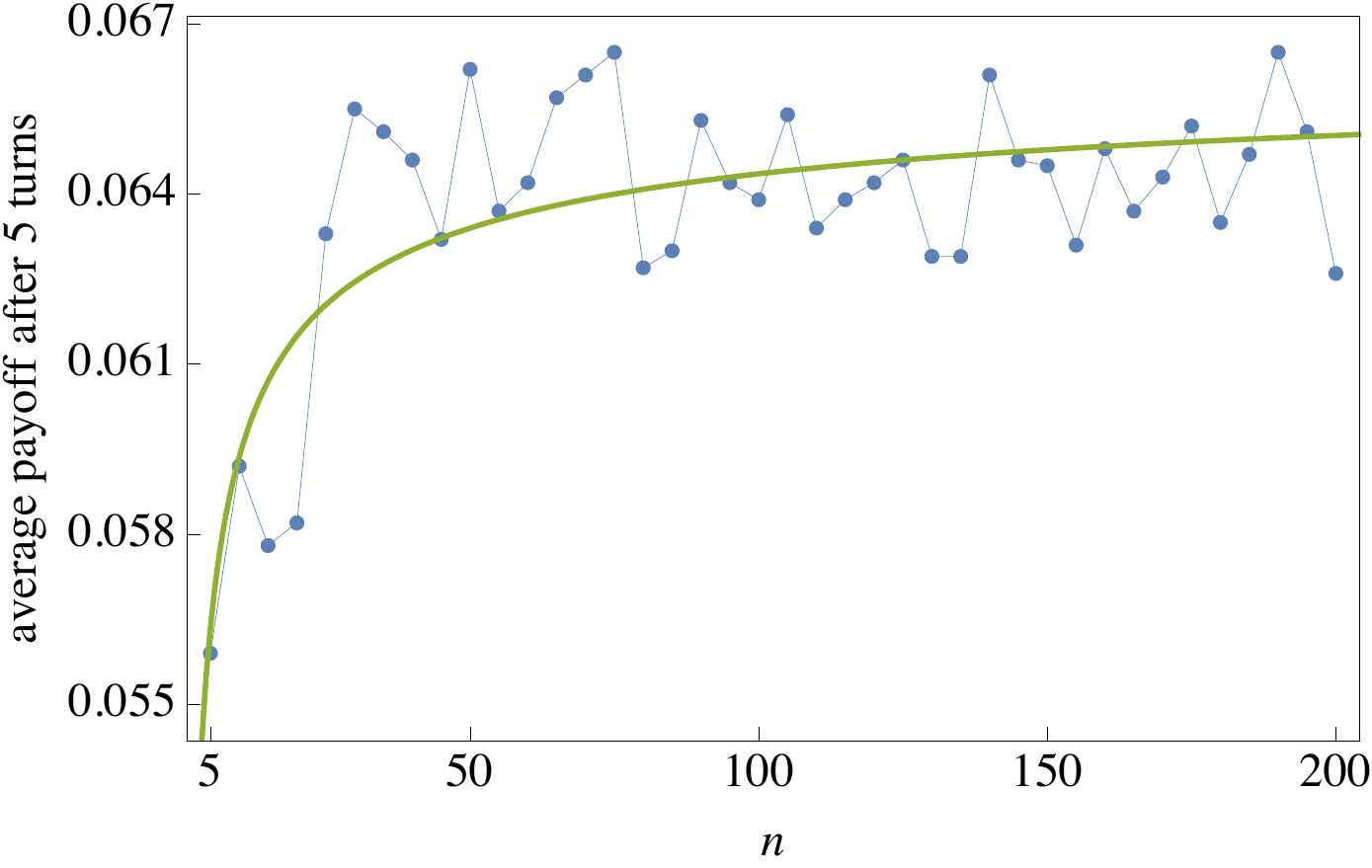}
\caption{Average score of \textsf{PREDICTOR} after $n$ turns in the tournament and evaluated every 5 turns (dots); also displayed is a fit to a $1/\sqrt{n}$ function.}
\label{fig:score_overview}
\end{figure}

\begin{figure}
\includegraphics[scale=0.758]{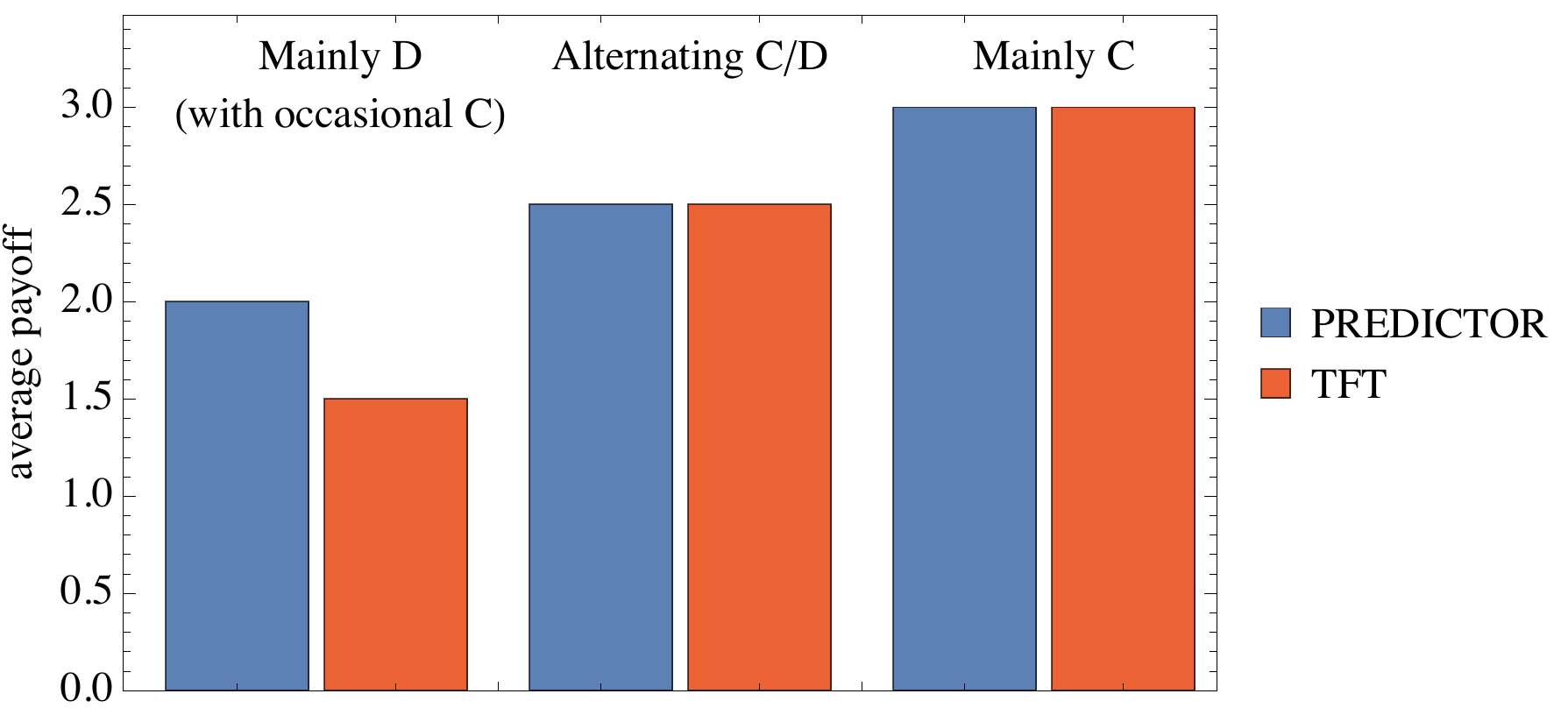}
\caption{Payoffs for a fictional match between \textsf{PREDICTOR} and \textsf{TFT}. Initially \textsf{PREDICTOR} chooses to always defect (with occasional cooperation), followed by an intermediate phase of alternation between cooperation and defection. Finally \textsf{PREDICTOR} settles on choosing sustained cooperation.}
\label{fig:score_TFT}
\end{figure}

\begin{figure}
\includegraphics[scale=0.758]{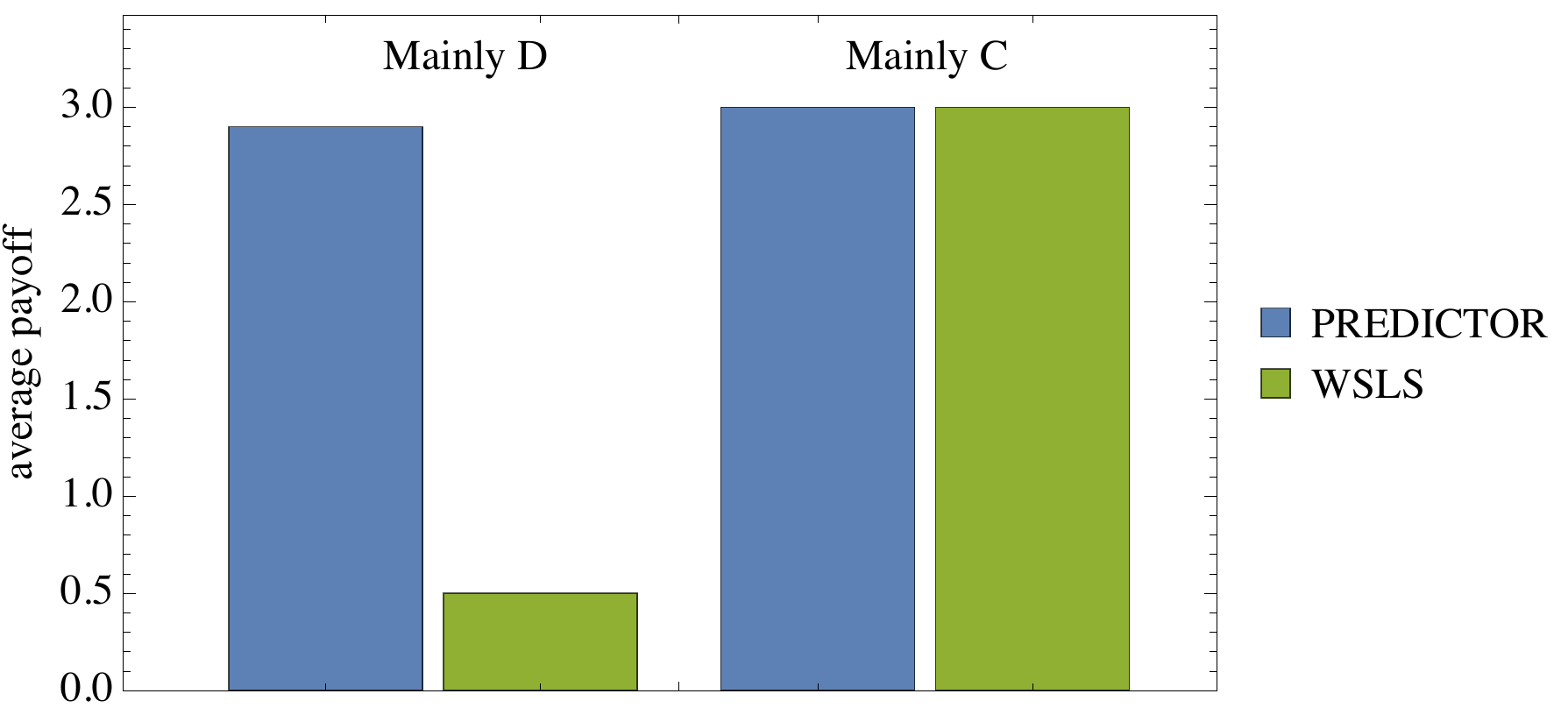}
\caption{Payoffs for a fictional match between \textsf{PREDICTOR} and \textsf{WSLS}. Initially \textsf{PREDICTOR} chooses to always defect (with occasional cooperation), leading to a good payoff with very low payoff for \textsf{WSLS}. Finally \textsf{PREDICTOR} settles on choosing sustained cooperation which brings an even (slightly) better payoff. }
\label{fig:score_WSLS}
\end{figure}

The evolution of average payoff for \textsf{PREDICTOR} during the tournament is shown in Fig. \ref{fig:score_overview}; payoff against selected matches in Figs. \ref{fig:score_TFT} and \ref{fig:score_WSLS} (corresponding tables can be found in the \textcolor{blue}{supporting information S4}).
\begin{table}[hbt]
\centering
\caption{Scoring in the round robin tournament; $N_\textrm{iter} = 5$, $N_\textrm{turn} = 200$, $p_\textrm{exp} = 0.1$. \textsf{PREDICTOR} comes first.}
\label{tab:res}
\begin{tabular}{lllll}
\\\toprule
place & strategy name & average payoff & standard error & $\#$ wins\\\hline
1. & \textsf{PREDICTOR} & 2.547 & 0.020 & 5 \\
2. & \textsf{GTFT} & 2.494 & 0.016 & 0\\
3. & \textsf{ZD-GTFT-2} & 2.432  & 0.024 & 2 \\
4. & \textsf{WSLS} & 2.418 & 0.015 & 0 \\
5. & \textsf{TFT} & 2.348 & 0.022 & 2 \\
6. & \textsf{ALLC} & 2.164 & 0.016 & 0 \\
7. & \textsf{RANDOM} & 2.118 & 0.050 & 5 \\
8. & \textsf{JOSS} & 2.075 & 0.032 & 7 \\
9. & \textsf{ALLD} & 2.058 & 0.018 & 7 \\
10. & \textsf{ZD-EXTORT-2} & 1.906 & 0.046 & 7 \\
\bottomrule
\end{tabular}
\end{table}

\subsection{Exploration vs Score}
\begin{figure}
\includegraphics[scale=0.8]{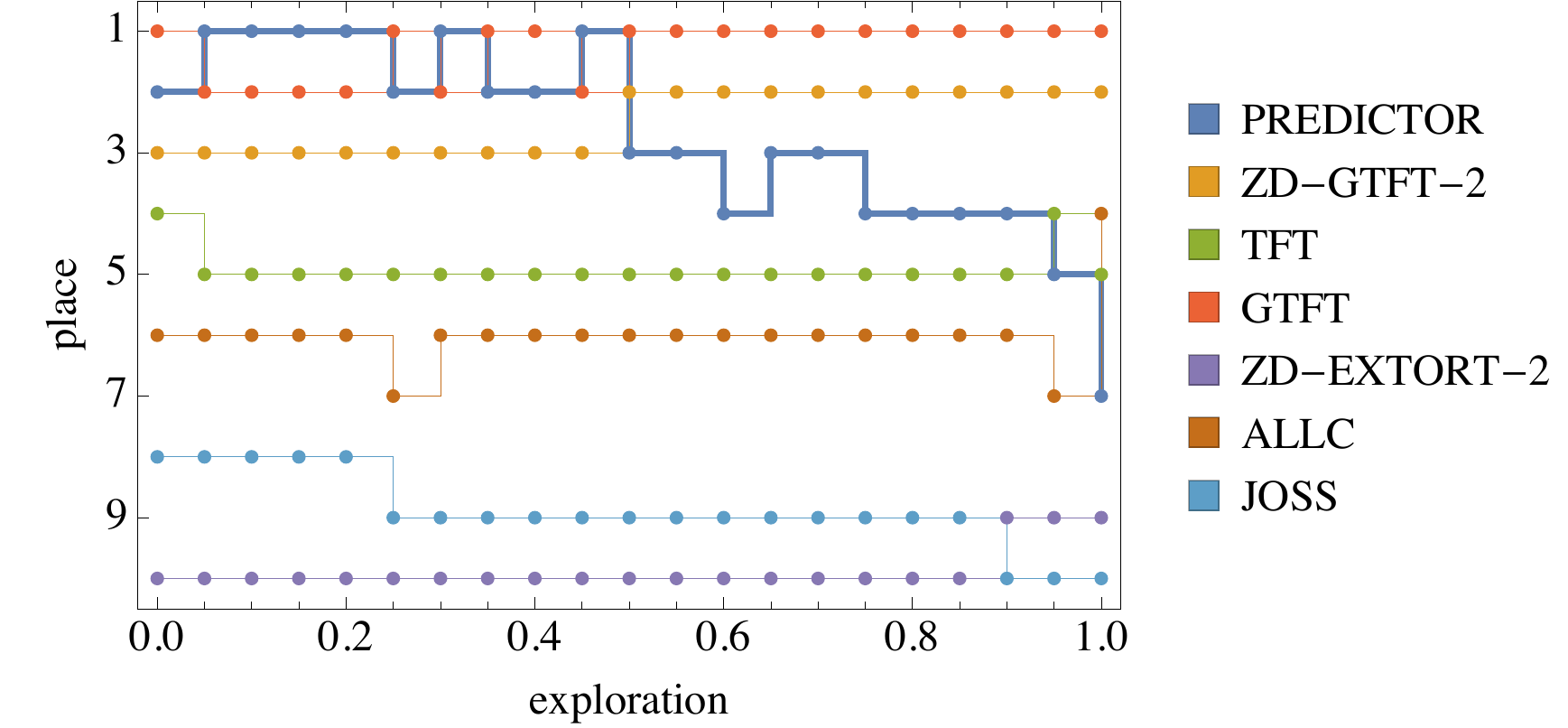}
\caption{Placement in the round robin as function of exploration (thick curve: \textsf{PREDICTOR}; thin curves: various other strategies). }
\label{fig:exploration}
\end{figure}

\begin{figure}
\includegraphics[scale=0.8]{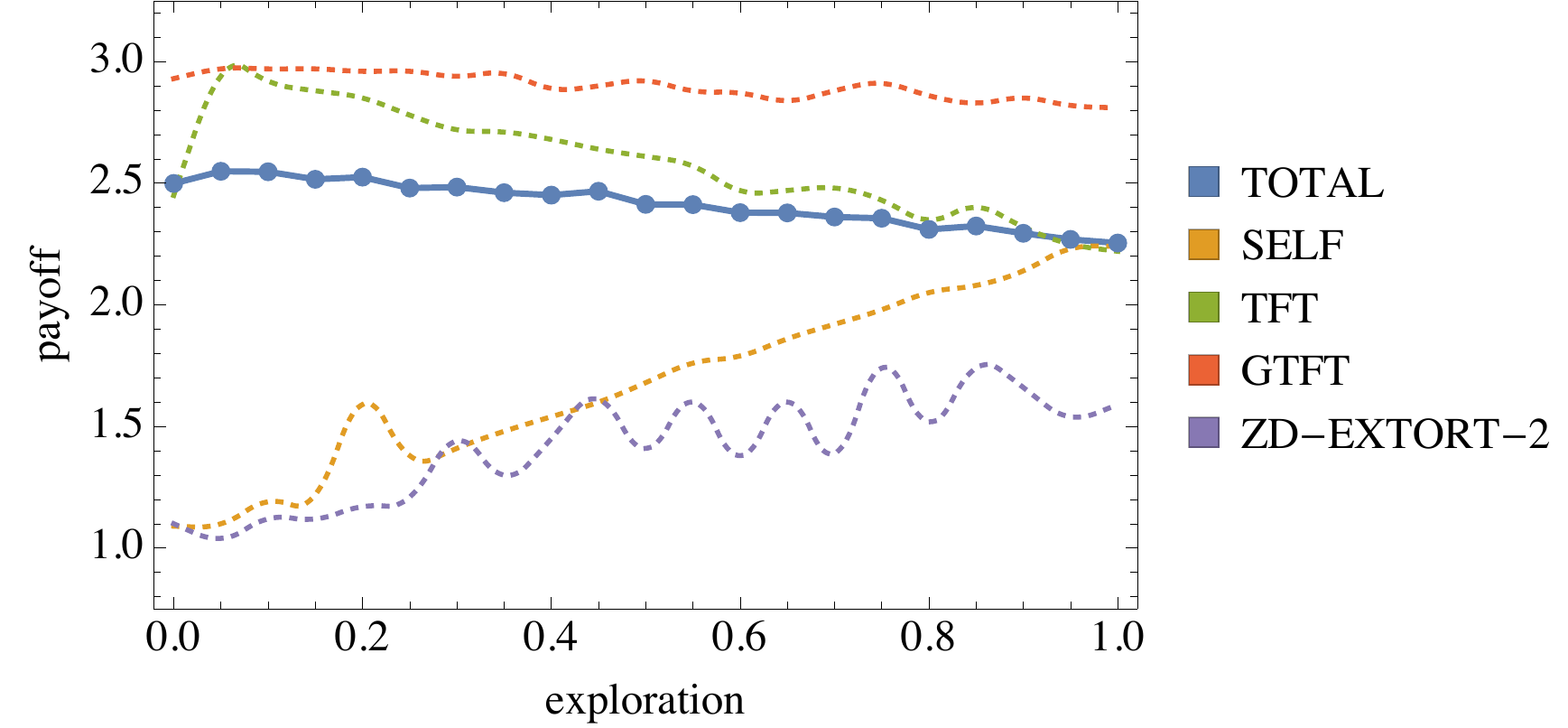}
\caption{Total payoff of \textsf{PREDICTOR} as function of exploration (thick curve; no interpolation used) and versus some specific strategies (thin curves; interpolation used). }
\label{fig:exploration2}
\end{figure}

Fig. \ref{fig:exploration} shows placement in the tournament as a function of different exploration parameters $p_\textrm{exp}$. Associated changes in average payoff of \textsf{PREDICTOR} when playing against selected strategies is shown in Fig. \ref{fig:exploration2}. (A corresponding table can be found in the \textcolor{blue}{supporting information S5}).

\clearpage
\section{Discussion}
\subsection{PREDICTOR learns from its experience}
In the simulated tournament, \textsf{PREDICTOR} was able to score the highest average payoff across all matches (Table \ref{tab:res})

\textsf{PREDICTOR} learned from its experience to make better decisions, based on an improved internal model. This is illustrated in Fig. \ref{fig:score_overview} where it is shown that \textsf{PREDICTOR}'s average score, evaluated every $\Delta n = 5$ turns, increases during the first $30-50$ turns of the game, enough to catapult \textsf{PREDICTOR} from a mediocre place in the competition to the top. The most important factor is the ability of \textsf{PREDICTOR} to improve its model during a match. This leads to better decisions which benefit \textsf{PREDICTOR} further on in the game.

Particular examples of how \text{PREDICTOR} updates its model and adjusts its strategy are shown in Figs. \ref{fig:score_TFT} and \ref{fig:score_WSLS}, where this is shown for its matches against \textsf{TFT} and \textsf{WSLS}, respectively. Initially, given maximal uncertainty about its opponent, \textsf{PREDICTOR} would choose defection. In the case of playing against \textsf{TFT} this would trigger retaliation on the next turn. Average payoff during this phase would be low for a number of turns. \textsf{PREDICTOR} then quickly (in a few turns) learns that \textsf{TFT}'s strategy is to always retaliate after initial defection. \textsf{PREDICTOR} would accordingly change its strategy and occasionally offer cooperation to \textsf{TFT}, followed by defection. Average payoff increases to $2.5$. Given enough experience, \textsf{PREDICTOR} changes its strategy to offer cooperation on every tun with an average payoff of $3$ which corresponds to the maximal achievable mutual payoff of $3+3$.
%

Even when playing against similar strategies which occasional defect, 
\textsf{PREDICTOR} would still offer cooperation, leading to larger payoff scores in the long run. So \textsf{PREDICTOR} might be characterized as ``conditionally forgiving'' (but it would keep retaliating when it thinks that the opponent is likely to $D$). 

When looking at the match against \textsf{WSLS}, one sees that \textsf{PREDICTOR} starts with sustained defection, reaping a total payoff of slightly below $3$ (due to occasional exploration). The reason for \textsf{PREDICTOR}'s comparatively high initial payoff is the fact that \textsf{WSLS} repeatedly alternates between cooperation and defection  -- according to the strategy of choosing a different action every time it encountered an adverse situation previously (either $DC$ or $DD$ leading to defection and cooperation in the next round respectively). 
%
After some time has elapsed in the game, \textsf{PREDICTOR} will change its strategy towards offering cooperation at every turn, which guarantees a maximal achievable mutual payoff. (When faced with such a situation, \textsf{WSLS} would never defect, except by mistake.)

We also find that the average number of matches won is not a good indicator for overall performance. Two strategies, \textsf{PREDICTOR} and \textsf{RANDOM}, won half of their matches, with \textsf{PREDICTOR} winning the tournament but \textsf{RANDOM} only achieving place seven. Also note that some strategies  (\textsf{GTFT} and \textsf{WSLS}) did not win any of their matches, whereas others which are similarly placed in the tournament (\textsf{ZD-GTFT-2} and \textsf{TFT}) won two out of ten. Tentatively, those strategies which are hard to beat come at the bottom of the tournament, because the price for their recalcitrance is the low payoff that comes with frequent defection.

Finally, and as already been noted by Axelrod after his first tournament \cite{Axe80a}, some strategies in the game might be low in overall score but act as ``kingmakers'', because the payoff coming from a match against these strategies makes difference between strategies which would (approximately) be equal otherwise (cf. the full payoff matrix in Table \ref{tab:po} in the \textcolor{blue}{supporting information S4}). 
One such strategy is \textsf{ALLC}, which always cooperates. Strategies such as \textsf{GTFT},  \textsf{ZD-GTFT-2} or \textsf{WSLS}, which came in at places 2-4, each reap an average payoff of $3$ from these matches, coming from sustained mutual cooperation. \textsf{PREDICTOR}, on the other hand, finds that repeated defection is the best strategy when playing against \textsf{ALLC}, thus being able to reap a payoff of approximately $5$.
Another kingmaker in the tournament is \textsf{JOSS} which plays similarly as \textsf{TFT} but with a small probability of defecting. While \textsf{ZD-GTFT-2} and \textsf{WSLS} reap an average payoff of ca. $2$, \textsf{PREDICTOR} and \textsf{GTFT} could reap an average payoff of $2.6$ and $2.5$ respectively, thus placing them slightly above their competitors in total score. \textsf{PREDICTOR} plays similarly against \textsf{JOSS} as it plays against \textsf{TFT}: first defecting, then learning to cooperate. However, \textsf{PREDICTOR} receives less payoffs when paired with \textsf{JOSS} compared to the pairing with \textsf{TFT}, because \textsf{JOSS} tends to defect occasionally.

\subsection{PREDICTOR uses exploration to improve its model}
Based on \textsf{PREDICTOR}'s internal model, its best decision is to choose the action which maximizes payoff over the next $2$ turns. However, there is an important caveat to this. In order to build a better model of the opponent's strategy it is sometimes advisable not to follow the above recipe of ``exploitation'' but to try out random actions in order to build a better model which would eventually result in reaping more payoffs on the long term. This can be inferred from Figs. \ref{fig:exploration} and \ref{fig:exploration2} (thick curves) where the effect of exploration on \textsf{PREDICTOR}'s placement in the tournament and its payoff against various other strategies is shown. 

In the specific examples discussed previously, the increase in average payoff towards the end of the game resulted mainly from exploration. 
Another occasion where an explorational strategy is better than a  pure exploitation strategy, is when \textsf{PREDICTOR} plays against itself. Given no exploration, \textsf{PREDICTOR} will always always choose to defect, resulting in a low mutual payoff of $1$. On the other hand, if \textsf{PREDICTOR} played a fully explorational strategy with $p_\textrm{exp} = 1.$, this would on average result in a payoff of $1/4\cdot (PO(CC)+PO(CD)+PO(DC)+P(DD)) = 2.25$. Intermediate exploration time would yield a payoff somewhere in between these extremes (Fig. \ref{fig:exploration2}; dotted curve labeled ``SELF'').

However, too much exploitation is not optimal since then \textsf{PREDICTOR} would not play well against some other strategies. (In the worst case, it would play as badly as a fully random strategy.) This indicates a trade-off between exploration and exploitation to be necessary to successfully compete in the tournament.

Interestingly, changing exploration time might have a big impact on the total average payoff (and thus the placement) but not on the number of winnings which stay roughly constant for all parameter choices (see Table \ref{tab:exploration} in the \textcolor{blue}{supporting information S5}). This is consistent with the idea that the number of matches won is irrelevant for determining the placing within a tournament. 
\subsection{PREDICTOR can evolve moral behavior}

Since the early tournaments of Axelrod \cite{Axe80a}, cooperating strategies tended to come up high. While ``defection'' is the unique Nash equilibrium of the single-shot game, it pays off to cooperate in the long run. This suggests that the \textsf{IDP} is a good model for the evolution of cooperation.
In our tournament strategies prone to cooperate came high in too (with the exception of \textsf{ALLC}) -- only surpassed by \textsf{PREDICTOR}. However, and in contrast to the original tournaments, \textsf{PREDICTOR} is not inherently ``nice'' or prone to cooperation but evolves such a behavior throughout the game.

It is instructive here to see what would happen when a strategy plays against itself. First, look at the example of the self-play of a retaliating but nice strategy, such as \textsf{TFT}. 
If \textsf{TFT} were initialized with a random action, and since \textsf{TFT} is a retaliating strategy, self-play would potentially result in constant defection throughout the iterated game. This means that under random initial conditions, \textsf{TFT} might perform badly when pitted against itself or a similar strategy. More precisely, averaging over all four possible initial actions, the average payoff for \textsf{TFT} when playing against itself would be $1/4 (3 + 1 + 2.5 + 2.5)  =  2.25$ (corresponding to initial pairings $CC$, $DD$, $CD$ and $DC$ respectively).  
For this and other reasons, the initial action of \textsf{TFT} is ``hardwired'' to cooperation in most implementations. However, this particular choice might create a different artifact in turn: many strategies will mutually cooperate until the end of the game and reap the maximal joint payoff of $3+3$ per turn.
Indeed, half of the strategies entered into the first tournament of Axelrod \cite{Axe80a} were of this type. The tendency for mutual cooperation as well as initially offering to cooperate, and hence the evolution of ``moral behavior'' is artificially ``steered''.

By contrast, \textsf{PREDICTOR} takes a random initial action and would first choose to defect under maximum uncertainty (i.e. when all model probabilities are set to $1/2$). So \textsf{PREDICTOR} is neither initially ``nice'' nor is it prone to cooperate \textit{per se}. As discussed previously, self-play of \textsf{PREDICTOR} would result in a low payoff of $1$ without exploration, and possibly increased by exploration to $2.25$, being much lower than the payoff from self-play of strategies that are ``hardwired'' to cooperate.

%

However, \textsf{PREDICTOR} is able to ``convince'' another strategy to cooperate (if it is in principle willing to) through offering cooperation at the cost of lower payoff in the present round for achieving ``the greater good'' in the long run. If, on the other hand, mutual cooperation were bad overall (which could be easily implemented by adjusting the payoff values of the stage game accordingly), \textsf{PREDICTOR} would not initiate mutual cooperation. What looks as if it was a ``morally good'' behavior which benefited every party, follows from \textsf{PREDICTOR}'s (selfish) assessment of the future consequences of its present actions.

Much in the spirit of the initial studies by Axelrod \cite{Axe84}, this shows how apparent morality could have emerged from ``intelligent selfishness''. More generally, this illustrates a basic virtue of good agent-based-models in the social and psychological sciences \cite{Jac17}:  While a complex behavior based on an explicit representation of ``perceived intention'' could be straightforwardly implemented –- for example by implementing a mechanism akin to a ``theory of mind'' (as suggested by Press \& Dyson \cite{Pre12}) -- ideally, a model would evolve a behavior that looks as if it stemmed from such a representation but in fact \textit{emerged} from a much simpler underlying mechanism -- such as the learning mechanism of \textsf{PREDICTOR}.
\section{Conclusion}
We have outlined \textsf{PREDICTOR}, a strategy that learns from its experience to choose optimal actions by modeling its opponent and predicting a (fictive) future. It has been shown that \textsf{PREDICTOR} is an efficient strategy for playing the iterated prisoner's dilemma, which is simple to implement. In the simulated tournament studied in this work, it achieved high average scores and won the tournament for various parameter settings. \textsf{PREDICTOR} relies on a brief phase of exploration to improve its model, and it is able to evolve morality from intrinsically selfish behavior.

\clearpage

\section*{Supporting Information}
\subsection*{S1: Memory-1 strategies selected for the tournament}

\begin{sidewaystable}[]
\centering
\begin{tabular}{l|llll|p{0.55\textwidth}}
\multicolumn{6}{c}{List of all strategies}\\
\toprule
Name: & $p(C|CC)$ & $p(C|CD)$ & $p(C|DC)$ & $pC|DD)$ & comments:\\\hline
\textsf{TFT} & 1. & 0. & 1. &  0. & cooperates if opponent previously cooperated, else retaliates\\
\textsf{GTFT} & 1. & 1/3 & 1. & 1/3 & like \textsf{TFT} but "forgives" with a probability of 1/3\\ 
\textsf{WSLS} & 1., & 0. & 0. & 1. & switches when payoff is less than 3, otherwise stays\\ 
\textsf{ALLD} & 0. & 0. & 0. & 0. & always defects\\
\textsf{ALLC} & 1. & 1. & 1. & 1. & always cooperates\\
\textsf{JOSS} & 9/10 & 0. & 9/10 & 0. & Like \textsf{TFT} but defects with probability 1/10\\
\textsf{ZDGTFT-2} & 1. & 1/8 & 1. & 1/4 & \textsf{ZD}-strategy; mild extortion\\ 
\textsf{ZDEXTORT-2} & 8/9 & 1/2 & 1/3 & 0 & \textsf{ZD}-strategy; strong extortion\\
\textsf{RANDOM} & 1/2 & 1/2 & 1/2 & 1/2 & always takes random action\\
\bottomrule
\end{tabular}
\caption{Overview over the memory-1 strategies used in Stewart \& Plotkin (2012)., including the top-3 performing strategies Tit-for-Tat (\textsf{TFT}), generous Tit-for-Tat (\textsf{GTFT}) and the \textsf{ZD}-strategy \textsf{ZD-GTFT}. \textsf{TFT, GTFT, WSLS, ALLC} and \textsf{JOSS} are ``nice" strategies that initially cooperate.}
\end{sidewaystable}

\clearpage
\subsection*{S2: Calculating Expected Payoffs}
\textsf{PREDICTOR's} decision affects (i) present payoffs but also (ii) future payoffs. We assume that \textsf{PREDICTOR} is equally weighting both (i.e. it makes no further assumption about a discount for future payoffs).

%
For \textit{every} (present or future) turn, \textsf{PREDICTOR} needs to estimate the payoff for what is likely to happen in case it chooses a particular action now + what will likely happen after he has chosen this action (up to the $n$-th turn). We will write $\langle PO \rangle (A_1\,A_2\hdots A_n)$ as the expected payoff of some specific sequence of choices of actions, $A_1\,A_2\hdots A_n$. At every turn, \textsf{PREDICTOR} is free to chose a new action. It therefore needs to estimate following distinct alternative sequences of actions:
\begin{equation}
\langle PO \rangle := \langle PO \rangle (A_1) + \langle PO \rangle (A_1 A_2) + \langle PO \rangle (A_1 A_2\hdots A_n)
\end{equation}
For example, setting $n=2$ and having $A_j \in \{C,D\}$, there are four alternatives ($k$ can take $2^n$ possible values, each corresponding to one action sequence):
\begin{eqnarray}
\langle PO \rangle_0 &=& \langle PO \rangle (C) + \langle PO \rangle (C,C)\\
\langle PO \rangle_1 &=& \langle PO \rangle (C) + \langle PO \rangle (C,D)\\
\langle PO \rangle_2 &=& \langle PO \rangle (D) + \langle PO \rangle (D,C)\\
\langle PO \rangle_3 &=& \langle PO \rangle (D) + \langle PO \rangle (D,D)
\end{eqnarray}
\textsf{PREDICTOR} evaluates these alternatives and then chooses the action $A_1 = C\ \textrm{or}\ D$ that corresponds to the largest $\langle PO \rangle_k$. \\

\noindent Expected payoffs for one turn can be calculated as:
\begin{equation}
\langle PO \rangle (A_1) = \sum_{A'_1} PO(A_1 A_1') p(A_1'|A_1 X_0) = \sum_{A'_1}PO(A_1 A'_1) p(A_1'|X_0),
\end{equation}
with  $X_0 = A_0 A'_0$ denoting the initially given action-pair. The last equality holds because the opponent's choice of $A_1'$ is independent from the player's choice $A_1$. We assume that \textsf{PREDICTOR} knows all possible payoffs, $PO(A_t A'_t)$; the conditional probabilities of the type $p(A'_t|A_{t-1} A'_{t-1})$ are contents of its model.
In our example from above, \textsf{PREDICTOR} would thus need to calculate 
\begin{equation}
\label{9}
\langle PO \rangle (C) = PO(CC) p(C|X_0) +  PO(CD) p(D|X_0), 
\end{equation}
and mutatis mutandis for $\langle PO \rangle (D)$.\\

\noindent For the next turn, \textsf{PREDICTOR} would need to calculate objects of the form:
%
\begin{eqnarray}
\langle PO \rangle (A_1,A_2) = \sum_{A'_2} PO(A_2\,A_2') \sum_{A'_1} p(A'_2|A'_1 A_1 X_0) p(A'_1|A_1 X_0)
\end{eqnarray}
In case the opponent is playing a memory-1 strategy, conditional probabilities only depend on the very last time step $p(A'_2|A_1 A'_1 X_0) = p(A'_2|A_1 A'_1)$. 
In general, this assumption is not true for all strategies, but given the property of memory-1 strategies discovered by Press \& Dyson (2012), this will be a good approximation throughout. The above expression would thus simplify:
\begin{equation}
\langle PO \rangle (A_1,A_2) = \sum_{A'_1, A'_2} PO(A_2\,A_2') p(A'_2|A'_1 A_1) p(A'_1|X_0)
\end{equation}
Again for the concrete example, one object to be evaluated would be:
%
\begin{eqnarray}
\langle PO \rangle (C,D) &=& \sum_{A'_1,A'_2} PO( D A_2' )  p(A'_2|C A'_1)p(A'_1|X_0)\\
&=& PO (D C) \left[ p(C|CC)p(C|X_0) + p(C|CD)p(D|X_0)\right]\\
&&+\,PO (D D) \left[ p(D|CC)p(C|X_0) + p(D|CD)p(D|X_0)\right].
\label{12} 
\end{eqnarray}
Note that, under the assumption of memory-1 strategies, no objects with $n\ge3$ need to be calculated, since $\langle PO \rangle (A_1 A_2 \hdots A_n) = \langle PO \rangle (A_{n-1} A_n)$. This means that, when playing against a memory-1 strategy, present choices can be ``undone'' with the next future choice. For the case of the \textsf{IPD}, all entries proportional to $PO(CD)$ will be zero, as well as many conditional probabilities (e.g. $p(C|DX) = 0$ for \textsf{TFT}). 

In the \textsf{IPD}, a game with a unique Nash equilibrium, the optimal choice at the last turn is known to be $D$, no matter what the opponent plays. This simplifies concrete implementation, and only $\langle PO \rangle_1$ and $\langle PO \rangle_3$ will need to be considered.
\subsection*{S3: Workflow of \textsf{PREDICTOR}}
\noindent Algorithm for PREDICTOR:
\begin{enumerate}
\item Initialize with random model
\item Repeat until game is finished
\begin{itemize} 
	\item Predict future courses (depth=2) of the game based on current model and previous actions
	\item Calculate expected payoffs for all possible futures (depth=2) and choose action for the next step that maximizes expected payoff 
	\item Record opponent's actions, receive payoff and update model
\end{itemize}
\item After each match: set model back model to random
\end{enumerate}

\subsection*{S4: Payoff-data for the round robin}
Table \ref{tab:po} lists all payoffs during the round robin tournament with parameters $N_\textrm{turn} = 200$, $N_\textrm{iter} = 5$ and $p_\textrm{exp} = 0.1$. Table \ref{tab:tft} list some specific turns for selected matches; parameters the same as above. 
\begin{sidewaystable}[]
\centering
\caption{Payoff-matrix for the round robin with parameter $N_\textrm{turn} = 200$, $N_\textrm{iter} = 5$ and $p_\textrm{exp} = 0.1$.}
\label{tab:po}
\begin{tabular}{lllll lllll l}
\\\toprule
& \textsf{PREDICTOR} & \textsf{RANDOM} & \textsf{ZD-GTFT-2} & \textsf{TFT} & \textsf{WSLS} & \textsf{ALLD} & \textsf{ZD-EXTORT-2} & \textsf{JOSS} &   \textsf{GTFT} & \textsf{ALLC} \\\hline
\textsf{PREDICTOR} & 1.188 & 2.971 & 2.946 & 2.917 & 2.906 & 0.954 & 1.125 & 2.625 & 2.969 & 4.866\\
\textsf{RANDOM}& 0.791 & 2.2785 & 2.489  & 2.221  & 2.296 & 0.515 & 1.701 & 2.057 & 2.837 & 3.994 \\
\textsf{ZD-GTFT-2} & 2.901 & 2.029 & 2.958&  2.918 & 2.945 & 0.755& 1.653& 2.192 &2.977& 2.994\\
\textsf{TFT} & 2.942 & 2.206 & 2.943 & 3.  &  3.  &  1.005 & 1.191& 1.192 & 3.  &  3.  \\ 
\textsf{WSLS} & 2.881 & 2.221 & 2.945 & 3.   & 3.  &  0.51  & 1.695 & 1.926 & 3.  &  3. \\
 \textsf{ALLD} & 1.184 &2.99  & 1.98 & 1.03 & 3.01 &  1.  &  1.   & 1.  &  2.408 & 5. \\
 \textsf{ZD-EXTORT-2} & 1.265 & 2.511 & 2.333 & 1.216 & 2.265 & 1.  &  1.146 & 1.085 & 2.754 & 3.49 \\
 \textsf{JOSS} & 3.09 &  2.332 & 2.607 & 1.217 & 2.091 & 1.  &  1.08  & 1.122 & 3.002 & 3.208 \\
\textsf{GTFT} & 2.859 & 1.957 & 2.962 & 3.  &  3. &   0.648 & 1.994 & 2.517 & 3. &   3.   \\
\textsf{ALLC} & 0.186 & 1.509 & 2.984 & 3. &   3.  &  0.005 & 2.265 2.688 & 3.  &  3. & \\   
\bottomrule
\end{tabular}
\end{sidewaystable}

\begin{table}[hbt]
\caption{\textsf{PREDICTOR} against \textsf{TFT} (top) and against \textsf{WSLS} (bottom) for the beginning of a match, with the model being most uncertain, and for the ending of a match, with the model being most informative. }
\label{tab:tft}
\centering
\hspace*{-1.5cm}
\begin{tabular}{p{0.15\textwidth} p{0.15\textwidth}  p{0.075\textwidth}| l p{0.15\textwidth} p{0.075\textwidth}| l p{0.15\textwidth} p{0.075\textwidth}}
\\\toprule
turn & \textsf{PREDICTOR} & \textsf{TFT} & turn & \textsf{PREDICTOR} & \textsf{TFT}  & turn & \textsf{PREDICTOR} & \textsf{TFT} \\\hline
1. & $D$ & $C$ & 11 & $D$ & $C$ & 191 & $C$ & $C$ \\
2. & $D$ & $D$ & 12 & $C$ & $D$ & 192 & $C$ & $C$ \\
3. & $D$ & $D$ & 13 & $D$ & $C$ & 193 & $C$ & $C$ \\
4. & $D$ & $D$ & 14 & $C$ & $D$ & 194 & $C$ & $C$ \\
5. & $D$ & $D$ & 15 & $D$ & $C$ & 195 & $C$ & $C$ \\
6. & $D$ & $D$ & 16 & $C$ & $D$ & 196 & $C$ & $C$ \\
7. & $C$ & $D$ & 17 & $D$ & $C$ & 197 & $C$ & $C$ \\
8. & $D$ & $C$ & 18 & $C$ & $D$ & 198 & $C$ & $C$ \\
9. & $C$ & $D$ & 19 & $D$ & $C$ & 199 & $C$ & $C$ \\
10. & $D$ & $C$ & 20 & $C$ & $D$ & 200 & $C$ & $C$ \\
\hline
$\langle PO \rangle_{n=10}$ & 2.0 & 1.5 && 2.5 & 2.5 && 3.0 & 3.0\\
\bottomrule\\
\end{tabular}
\begin{tabular}{p{0.15\textwidth} p{0.15\textwidth}  p{0.075\textwidth}| l p{0.15\textwidth} p{0.075\textwidth}}
\toprule
turn & \textsf{PREDICTOR} & \textsf{WSLS} & turn & \textsf{PREDICTOR} & \textsf{WSLS}   \\\hline
1. & $D$ & $C$ & 191 & $C$ & $C$  \\
2. & $D$ & $D$ & 192 & $C$ & $C$  \\
3. & $D$ & $C$ & 193 & $C$ & $C$  \\
4. & $D$ & $D$ & 194 & $C$ & $C$  \\
5. & $D$ & $C$ & 195 & $C$ & $C$  \\
6. & $D$ & $D$ & 196 & $C$ & $C$  \\
7. & $D$ & $C$ & 197 & $C$ & $C$  \\
8. & $D$ & $D$ & 198 & $C$ & $C$  \\
9. & $D$ & $C$ & 199 & $C$ & $C$  \\
10. & $D$ & $D$ & 200 & $C$ & $C$ \\
\hline
$\langle PO \rangle_{n=10}$ & 3.0 & 0.5 && 3 & 3 \\
\bottomrule
\end{tabular}
\end{table}
\clearpage
\subsection*{S5: Payoff as function of exploration}
\begin{table}[hbt]
\centering
\caption{Fraction of exploration, \textsf{PREDICTOR}'s average payoff, the difference to \textsf{ZD-GTFT-2}'s payoff, place in the tournament, number of wins (for selected values of $p_\textrm{exp})$.}
\label{tab:exploration}
\begin{tabular}{p{0.075\textwidth}llll}
\\\toprule
$p_\textrm{exp}$ & $\langle PO \rangle$ &  $\Delta_{ \textsf{ZD-GTFT-2}} $  & place & $\#$ wins
\\\hline
0.      &		2.499 		&		0.052		&		2 &	    5 \\
0.05  & 	2.549 		&      0.099 		& 		1 &    	4\\  
0.1  	 & 	2.547 		& 		0.115		& 		1 &    	5\\   
0.15  & 	2.516 		&      0.065 		&      1 &    	4\\
0.2   &		2.536 		& 		0.083		& 		1 &   	4\\   
0.25 & 		2.48 		& 	   		0.049	&	 	2 &    	5\\  
0.3   &		2.484 		&	   		0.049 	& 		1 &    	4\\
0.35 & 		2.4618 		& 	   		0.023 	& 		2 &    	5\\   
0.4   &		2.451 		&	   		0.027 	&   	2 &    	4\\
0.45 & 		2.467 		& 	   		0.034 	& 		1 &    	5\\   
0.5   & 		2.413       &  	-0.023  	&	  	3    &  	4\\   
0.55  & 	2.412    		&    	-0.011  	& 	  	3   &  	4\\  
0.6    &		2.379       & 		-0.027     & 	  	4  &   	5\\   
0.65  & 	2.378  		& 		-0.011  	&     	3  &   	4\\   
0.7   & 		2.361       &  	-0.044  	& 	  	3   &  	5\\   
0.75  & 	2.356  	    & 		-0.050   	&	  	4   &  	5\\   
0.8    &		2.310   	& 		-0.077    		& 	  	4    & 	4\\
0.85  & 	2.324  	    & 		-0.051   	& 	  	4   &  	4 \\
0.9    &		2.294  	    & 		-0.099  	&  	4  &   	5\\   
0.95  & 	2.269  	    & 		-0.119  	& 	  	5 &    	5\\  
1. &     		2.254  	    & 		-0.093		&		7 &    	4 \\
\bottomrule
\end{tabular}
\end{table}

\end{document}